\newif\ifdraft
\newif\ifcomments
\PassOptionsToPackage{prologue,dvipsnames}{xcolor}
\ifdraft
    \documentclass[anonymous,sigconf,nonacm]{acmart}
\else
    \documentclass[sigconf,nonacm]{acmart}
\fi

\usepackage[nameinlink,noabbrev]{cleveref}
\usepackage{chngcntr}

\usepackage[detect-all,group-separator={,}]{siunitx}
\usepackage{amsmath,amsfonts}
\usepackage{balance}
\usepackage{booktabs}
\usepackage{comment}
\usepackage{csquotes}
\usepackage{algorithmic}
\usepackage[inline]{enumitem}
\usepackage{natbib}
\usepackage{graphicx}
\usepackage{lipsum}
\usepackage{multicol}
\usepackage{multirow}
\usepackage{textcomp}
\usepackage[dvipsnames]{xcolor}
\usepackage[most]{tcolorbox} 

\usepackage{subcaption}
\usepackage{mathtools}
\usepackage{hyperref}
\usepackage[nameinlink]{cleveref}

\usepackage{chngcntr} 
\usepackage{caption}
\usepackage{flushend}
\newcommand{\Rhea}{Rhea}

\counterwithin{promptctr}{section}              

\newtcolorbox[
    auto counter, 
    use counter=promptctr, 
    number within=section, 
]{promptbox}[1][]{%
  enhanced,
  breakable,
  colback=cyan!5!white,
  colframe=cyan!60!black,
  colbacktitle=cyan!30!white,
  coltitle=black,
  fonttitle=\bfseries,
  title=Prompt~\thetcbcounter, 
  list entry={Prompt~\thetcbcounter},
  rounded corners,
  boxrule=0.7pt,
  top=6pt, bottom=6pt, left=6pt, right=6pt,
  attach boxed title to top left={yshift=-2mm,xshift=2mm},
  boxed title style={rounded corners, boxrule=0pt, colframe=cyan!60!black},
  #1 
}

\crefname{promptctr}{Prompt}{Prompts}

\ifcomments
  \newcommand{\ian}[1]{{\textcolor{orange}{\textbf{Ian:}~\enquote{#1}}}}
  \newcommand{\kyle}[1]{{\textcolor{blue}{\textbf{Kyle:}~\enquote{#1}}}}
  \newcommand{\ryan}[1]{{\textcolor{cyan}{\textbf{Ryan:}~\enquote{#1}}}}
  \newcommand{\reid}[1]{{\textcolor{teal}{\textbf{Reid:}~\enquote{#1}}}}
  \newcommand{\maxime}[1]{{\textcolor{red}{\textbf{Maxime:}~\enquote{#1}}}}
  \newcommand{\tanjin}[1]{{\textcolor{magenta}{\textbf{Tanjin:}~\enquote{#1}}}}
  \newcommand{\haochen}[1]{{\textcolor{ForestGreen}
  {\textbf{Haochen:}~\enquote{#1}}}}
  
  \newcommand{\fm}[1]{{\textcolor{red}{\textbf{FIXME:}~\enquote{#1}}}}
\else
  \newcommand{\ian}[1]{}
  \newcommand{\kyle}[1]{}
  \newcommand{\ryan}[1]{}
  \newcommand{\reid}[1]{}
  \newcommand{\maxime}[1]{}
  \newcommand{\tanjin}[1]{}
  \newcommand{\haochen}[1]{}
  \newcommand{\fm}[1]{}
\fi

\begin{document}

\title{Experiences with Model Context Protocol Servers for Science and High Performance Computing}
\newcommand{\uchicago}{\textsuperscript{*}}
\newcommand{\argonne}{\textsuperscript{\textdagger}}
\newcommand{\uic}{\textsuperscript{\textdaggerdbl}}

\author{
    Haochen Pan\uchicago\argonne,
    Ryan Chard\argonne\uchicago,
    Reid Mello\uchicago,
    Christopher Grams\argonne\uic,
    Tanjin He\argonne,
}
\author{
    Alexander Brace\argonne\uchicago,
    Owen Price Skelly\uchicago,
    Will Engler\uchicago,
    Hayden Holbrook\uchicago,
    Song Young Oh\uchicago,
}
\author{
    Maxime Gonthier\uchicago\argonne,
    Michael Papka\argonne\uic,
    Ben Blaiszik\uchicago,
    Kyle Chard\argonne\uchicago,
    Ian Foster\argonne\uchicago
}

\affiliation[obeypunctuation=true]{\uchicago University of Chicago; Chicago, IL, \country{United States}}
\affiliation[obeypunctuation=true]{\argonne Argonne National Laboratory; Lemont, IL, \country{United States}}
\affiliation[obeypunctuation=true]{\uic University of Illinois Chicago; Chicago, IL, \country{United States}}
\renewcommand{\shortauthors}{Pan and Chard et al.}

\begin{abstract}

Large language model (LLM)-powered agents are increasingly used to plan and execute scientific workflows, yet most research cyberinfrastructure (CI) exposes heterogeneous APIs and implements security models that present barriers for use by agents. We report on our experience using the Model Context Protocol (MCP) as a unifying interface that makes research capabilities discoverable, invokable, and composable. Our approach is pragmatic: we implement thin MCP servers over mature services, including Globus Transfer, Compute, and Search; status APIs exposed by computing facilities; Octopus event fabric; and domain-specific tools such as Garden and Galaxy. We use case studies in computational chemistry, bioinformatics, quantum chemistry, and filesystem monitoring to illustrate how this MCP-oriented architecture can be used in practice. 
We distill lessons learned and outline open challenges in evaluation and trust for agent-led science.

\end{abstract}

\maketitle
\section{Introduction}

Generative AI is reshaping how scientists design, run, and manage computational experiments. Rapid development of reasoning models~\cite{jaech2024openai,guo2025deepseek} and agentic libraries~\cite{langchain,pauloski2025empowering} encourage us to envision the scientific process being driven by AI agents, capable of planning multi-step analyses, designing and running experiments, and coordinating workflows across distributed facilities. 
However, realizing this vision in practice requires overcoming the inherent heterogeneity of research cyberinfrastructure (CI), where each computer, service, instrument, tool, and database has its own APIs, security models, and operational requirements. 

We explore the Model Context Protocol (MCP) as a unifying interface to empower AI agents to discover, invoke, and coordinate capabilities for scientific workloads. 
MCP defines simple, typed interfaces for tools and resources, plus feedback channels that let agents observe progress and recover from errors. 
Here, we implement MCP servers for three mature research services---Globus Transfer~\cite{ananthakrishnan2015globus}, Compute~\cite{funcX2020chard}, and Search~\cite{Ananthakrishnan2018Globus}; facility status endpoints at the Argonne Leadership Computing Facility (ALCF) and National Energy Research Scientific Computing Center (NERSC); Octopus event fabric~\cite{pan_octopus_2024}; and the Garden~\cite{garden-ai} and \Rhea~\cite{Rhea} domain ecosystems. 
The full collection of MCP servers will be available soon at our GitHub repository~\cite{science_mcps}.
These servers provide agents with capabilities needed to perform distributed scientific computing at scale, for example by enabling secure, high-performance data transfer between storage systems and execution of Python functions on remote computers. 

We evaluate this MCP-oriented architecture through four classes of workflows: (i) computational chemistry model discovery and inference; (ii) multi-site bioinformatics pipelines; (iii) quantum chemistry with federated computations; and (iv) filesystem monitoring and usage evaluation. 
In each case, we show how an AI agent can orchestrate complex, and even multi-site workflows, by invoking and dynamically generating the necessary functions to leverage heterogeneous resources.

Our contributions include investigating MCP-oriented architecture for scientific computing and research CI, developing reference MCP implementations, and distilling lessons on reliably connecting HPC and scientific services to agentic workflows. 
We observe that it is beneficial to build thin MCP adapters for broad research services 
rather than to create new services, and to separate discovery from invocation (e.g., in~\Rhea) to manage large tool ecosystems, and that agents can remove the need to create custom glue code to combine scientific applications---instead generating it as needed. 
We also note challenges, including hosting MCP services for research CI that span administrative domains under existing authentication models, evaluating the reliability of agent-driven workflows, and strengthening resilience and recovery for long-running tasks.

\section{Background \& Related Work}

We first describe MCP and its use in science, and then review other methods used to provide agents with access to tools. 

\subsection{Model Context Protocol}

MCP, proposed in 2024, is an open interface specification for connecting agents to data sources and tools~\cite{anthropic_mcp}.
It defines three core primitives: 
\textit{resources}, read-only data sources from which an agent can gather information, such as a file, database records, or system status feed; 
\textit{tools}, capabilities that an agent can invoke to perform an action, such as executing a computational analysis or transferring a file; and 
\textit{prompts}, reusable templates that can be used to generate contextualized prompts for an agent, such as an HPC batch submission template that can be parameterized with job name and number of required nodes.
MCP allows agents to discover available resources, tools, and prompts; use tools; and incorporate structured results into reasoning processes.
Here, we focus on 
\textit{resources} and 
\textit{tools} as two primary means of exposing research CI to agents. 

MCP has attracted broad interest, with more than 8000 open-source servers now listed in the Glama MCP directory~\cite{glama_mcp_servers}. 
MCP is also being adopted in scientific domains to connect tools and datasets to LLMs. 
For example, \textit{mcp.science} \cite{mcp_science_github} publishes MCP servers for various science applications, such as the Materials Project~\cite{next_gen_materials_project} to query and visualize materials data, GPAW~\cite{Mortensen2024GPAW} for density functional theory calculations, and neuroscience model analysis through the
NEMAD API~\cite{itani2024northeast}.
In biomedicine, \textit{Bio-Agents MCP}~\cite{bio_agents_mcp} provides MCP servers for the Protein Data Bank~\cite{berman2000pdb} and ChEMBL~\cite{gaulton2012chembl}.
Other MCP servers~\cite{open-search-mcp,arxiv-mcp-server} enable agents to search and retrieve scholarly content from arXiv, PubMed, bioRxiv, and medRxiv, returning structured metadata or full texts to support evidence gathering and experiment planning.  

\subsection{Other Methods for Agent Tool Use}

FutureHouse's AI agent PaperQA2~\cite{skarlinski2024language,paper-qa} generates Wikipedia-style summaries grounded in evidence extracted from scientific papers.
The agent orchestrates several tools to retrieve documents, extract relevant paragraphs, and synthesize answers with citations.

Biomni~\cite{Huang2025Biomni} couples an agentic planner 
with a curated execution environment. 
It leverages LangChain’s tool-calling protocol~\cite{langchain-tool-calling} to integrate 
150 specialized tools, 59 domain databases, and 105 software packages; 
the agent iteratively plans, executes code, inspects results, and updates its strategy until objectives are met.

Broadly, research on tool-using agents spans several fronts. Synthetic data pipelines are used to generate large-scale, high-quality data to train models to effectively use tools~\cite{tang2023toolalpaca,liu2025toolace,wang2025toolflow}. Well-structured API documentation can also lead to reliable tool use without specific training~\cite{hsieh2023tool}. The development of ecosystems helps connect models to vast APIs~\cite{liang2024taskmatrix} and can proactively assemble MCP toolchains to reduce context overhead~\cite{fei2025mcp}. Other efforts include benchmarks to stabilize and evaluate MCP tool use~\cite{guo2024stabletoolench, fan2025mcptoolbench,gao2025mcp}, reliability efforts to test and codify tool use~\cite{milev2025toolfuzz,ding2025toolcoder}, and security studies surface new risks and vulnerabilities~\cite{hasan2025model,li2025we}.

\section{Problem Definition}

When presented with a prompt, an AI agent (e.g., Claude Desktop) engages an LLM (e.g., Claude Sonnet 4) to develop and run an execution plan, including by making calls to MCP servers.

This 
plan constitutes an \textbf{agentic application} that is executed as a result of the supplied user prompt. 
Such an application executes within a particular \textbf{application context} that comprises the user prompt ($p$), the coordinating LLM ($L$), any user credentials ($\Phi_{\text{user}}$), and a set of MCP servers ($\mathcal{M}$), plus (optionally) a set of computing sites ($\mathcal{S}$) that can be accessed via MCP servers. 

An \textbf{MCP server} exposes a collection of capabilities and handles authorization requests.
Each server $M_k\in \mathcal{M}$ is defined as
    $M_k = \langle \mathcal{C}_k, \Phi_k \rangle$,
where $\mathcal{C}_k$ is the set of capabilities exposed by the server and $\Phi_k$ is the authentication and authorization client it employs (e.g., OAuth 2.0 client).
For \textit{discovery-enabled} servers, the set of capabilities is dynamic. A discovery call using a natural-language query $q$ can materialize additional capabilities, denoted $\mathcal{C}_k(q)$, which are added to the server's initially available set.
As noted above, an MCP server can also define resources ($\mathcal{X}$) and prompts ($\mathcal{P}$). For brevity, we omit these in the following, but a complete definition of an MCP server is a 4-tuple $\langle \mathcal{C}_k, \Phi_k, \mathcal{P}_k, \mathcal{X}_k \rangle$.

An MCP server \textbf{capability} is an invokable action defined by its interface (inputs and outputs), description, and execution requirements, $C_j = \langle \mathcal{I}_j, \mathcal{E}_j, \mathcal{D}_j, \mathcal{R}_j \rangle$, where $\mathcal{I}_j$ is required inputs, $\mathcal{E}_j$ expected outputs, $\mathcal{D}_j$ description (name and documentation), and $\mathcal{R}_j$ requirements on site software ($\Pi_i$) and resources ($\Sigma_i$).

Finally, a \textbf{computing site} is an execution environment, such as a specific supercomputer (e.g., Aurora or Polaris at ALCF). 
Each site $S_i\in \mathcal{S}$ is defined by its software and hardware,
    $S_i = \langle \Pi_i, \Sigma_i \rangle$,
where $\Pi_i$ is the set of installed software packages and $\Sigma_i$ is the set of available computational resources (CPUs, GPUs, etc.). 

Having defined these various terms, we now return to describing what happens during execution of an agentic application.
The core process is a three-step workflow in which the coordinating LLM $L$ converts a user prompt $p$ into a final output $\mathcal{O}$, as follows.

    \textbf{Plan:} The agent uses LLM $L$ to convert the user prompt $p$ into an abstract plan, $\mathcal{T}$: 
    a set of 
    high-level goals not yet tied to specific capabilities or sites:
        $\mathcal{T} = \textsf{Plan}(p, L)$.
    This stage \textit{succeeds} if the LLM generates a coherent, actionable abstract plan. 
    
    \textbf{Resolve:} The agent then translates the abstract plan $\mathcal{T}$ into a concrete plan $\mathcal{R}$ by finding a feasible tuple $(S_i, C_j, M_k)$ for each abstract task $t \in \mathcal{T}$: $\mathcal{R} = \textsf{Resolve}(\mathcal{T}, \mathcal{M}, \mathcal{S})$. $L$ evaluates each capability's interface, description, and execution requirements to select an appropriate $C_j$ for each task $t$.
    A tuple is 
    feasible if: 
    
    \begin{itemize}[left=1em]
        \item 
        The capability $C_j$ is available from server $M_k$. For discovery-enabled servers, the capability set of $M_k$ is materialized as $\mathcal{C}_k(q)$ using a query $q$ derived from an abstract task $t$.
                
        \item 
        The site $S_i$ satisfies all 
        technical 
        requirements $\mathcal{R}_j$ of the capability $C_j$: i.e., the required software and hardware are available in the site's definition, $\langle \Pi_i, \Sigma_i \rangle$.
                
    \end{itemize}

    This stage \textit{succeeds} if a feasible tuple is found for every abstract task in $\mathcal{T}$, resulting in a complete concrete plan $\mathcal{R}$.

    \textbf{Execute:} The agent executes the concrete plan $\mathcal{R}$ by processing each tuple $(S_i, C_j, M_k)$ in turn. Each step comprises two phases:
        \begin{enumerate}
            \item 
            The agent requests authorization using the server's client, $\Phi_k$, with the user's credentials, $\Phi_{\text{user}}$.
            
            \item 
            If authorization is granted, the agent invokes the capability $C_j$ on the target site $S_i$.
        \end{enumerate}
        
    This stage \textit{succeeds} if every step in $\mathcal{R}$ is authorized and invoked without error, yielding the final output $\mathcal{O} = \textsf{Execute}(\mathcal{R}, \Phi_{\text{user}})$.

The complete workflow, $\mathcal{W}(p)$, which can be expressed as the composition of these three stages, is considered successful if and only if all three stages complete successfully:
\begin{equation}
\boxed{
\mathcal{W}(p) = \textsf{Execute}\left(\textsf{Resolve}\left(\textsf{Plan}(p, L), \mathcal{M}, \mathcal{S}\right), \Phi_{\text{user}}\right)
}
\end{equation}

\section{Scientific MCP Servers}

We have implemented MCP servers for Globus Transfer, Compute, and Search services; facility status for ALCF and NERSC; the Octopus event fabric; the Garden platform; and~\Rhea. All servers are deployed as separate Docker containers using the \texttt{streamable-HTTP} transport, which supports multi-client, bidirectional communication and avoids the per-process overhead of the \texttt{stdio} transport. The use of containers also isolates SDK dependencies and allows credentials to be provided at runtime rather than hardcoded into images. Each container includes a minimal Python runtime, only the necessary service-specific packages, and a lightweight entrypoint script with optional server parameters. We describe these servers and their capabilities below.

The \textbf{Globus Transfer} MCP server
implements tools 
that agents can use to interact with the Globus Transfer service to discover collections, browse file systems, and transfer files between collections. 
The server handles authentication flows, manages transfer task lifecycles, and provides interactive status monitoring. 

The \textbf{Globus Compute} MCP server 
exposes tools for agents to execute 
Python and Shell functions
on remote endpoints, monitor function execution, and retrieve function results. 
It manages authentication flows to perform actions securely on the user's behalf. 

The \textbf{Globus Search} MCP server exposes tools for agents to create, delete, and list  Globus Search indexes, and to ingest, delete, and query records in specific indexes. 
Thus, it allows agents to discover datasets and research artifacts across distributed repositories, with conversational queries 
translated into structured search requests and 
results converted into structured responses.

The \textbf{Computing Facility} MCP server implements resources and tools that agents can use to obtain real-time operational status and resource availability of ALCF and NERSC computers. 
It exposes facility information as MCP resources 
representing the state of individual systems, and provides MCP tools 
for retrieving system health, queue status, maintenance schedules, and resource utilization. 
The server translates complex operational data into structured, queryable reports, enabling decision-making about where and when to submit computational tasks.

The \textbf{Octopus} MCP server equips agents with event streaming capabilities. 
Its backend, the Octopus event fabric, is a cloud-to-edge streaming platform built on AWS Managed Streaming for Kafka and secured with Globus Auth. Users can create and delete topics, update configurations, and truncate events within topics. Agents can publish or consume events directly under a user's identity.

The \textbf{Garden} MCP server provides tools for agents to discover and run
scientific machine learning models. 
The Garden platform catalogs a diverse collection of domain-specific models and, through the MCP interface, agents can 
discover and inspect published models, 
and invoke them to perform inference tasks on either the cloud or available HPC resources.

The \textbf{\Rhea}~MCP server provides agents with access to the many bioinformatics tools in the Galaxy Toolshed~\cite{galaxy-toolshed}. 
Handling many tools in MCP servers risks overwhelming the agent’s context window.
\Rhea~addresses this challenge by providing a dynamic interface. 
Specifically, the MCP server uses Retrieval-Augmented Generation (RAG)~\cite{lewis2020retrieval} over textual descriptions from Galaxy Toolshed documentation, help strings, and parameter schemas. 
These descriptions are embedded using \texttt{Qwen3-Embedding-0.6B}, and the MCP server exposes one tool, \texttt{find\_tools}, that accepts a natural-language query for a desired capability. 
When invoked, it performs RAG to identify the top-$k$ bioinformatics applications most relevant to the query in embedding space and then dynamically generates corresponding MCP tools. 
The MCP server uses the protocol’s notification channel mechanism to alert the agent of the new tools, prompting the agent to refresh its tool list and enabling the agent to invoke the new tools.

\section{Application Case Studies}

\begin{figure*}[t]
\centering

\begin{minipage}[b]{0.48\linewidth}
\centering
\includegraphics[width=\linewidth, page=2, trim={0.05cm, 8.75cm, 7.4cm, 0cm}, clip]{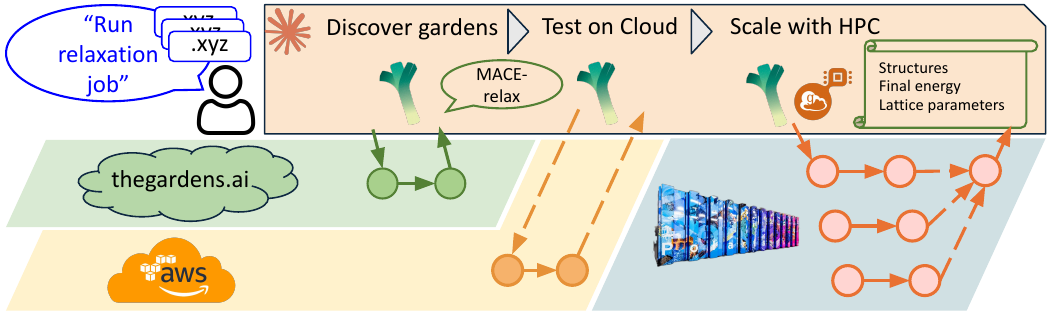}
\subcaption{The Garden workflow. The MACE model is discovered via the Garden MCP server and then inference is run on AWS or HPC.}
\label{fig:garden-app}
\end{minipage}
\hfill
\begin{minipage}[b]{0.48\linewidth}
\centering
\includegraphics[width=\linewidth, page=3, trim={0.05cm, 8.75cm, 7.4cm, 0}, clip]{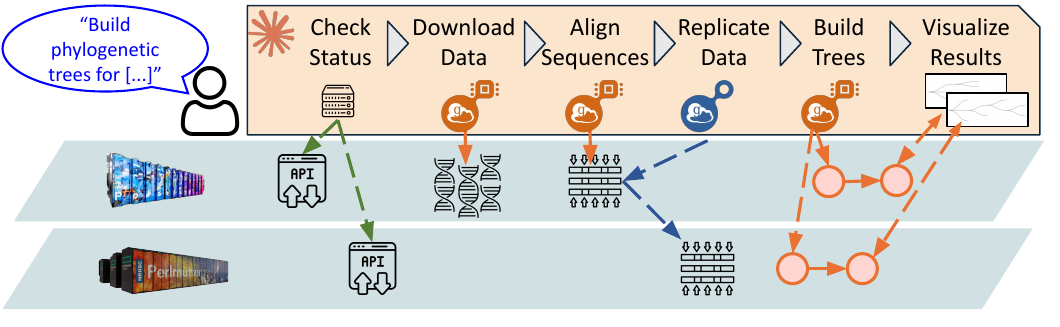}
\subcaption{The phylogenetic workflow. The facility MCP server determines resource availability before data are downloaded, aligned, and analyzed at ALCF and NERSC using the Globus Compute MCP server.}
\label{fig:bio-app}
\end{minipage}
\vspace{1em}

\begin{minipage}[b]{0.48\linewidth}
\centering
\includegraphics[width=\linewidth, page=4, trim={0, 10.5cm, 9cm, 0}, clip]{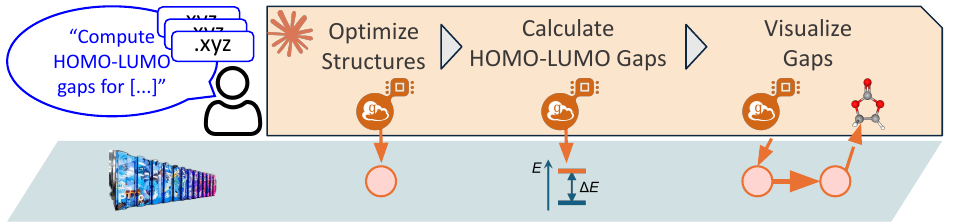}
\subcaption{The quantum chemistry workflow. The Globus Compute MCP server is used to optimize structures, calculate HOMO-LUMO gaps, and visualize results using ALCF's Polaris.}
\label{fig:chem-app}
\end{minipage}
\hfill
\begin{minipage}[b]{0.48\linewidth}
\centering
\includegraphics[width=\linewidth, page=5, trim={0, 10.5cm, 9cm, 0}, clip]{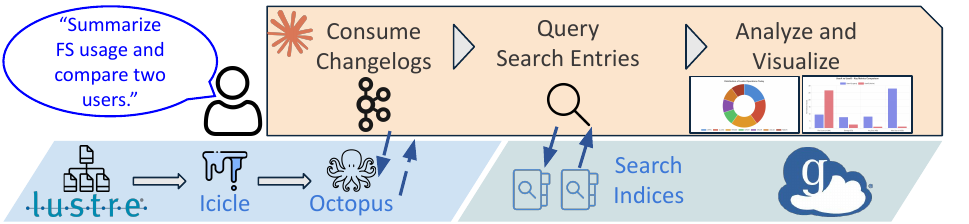}
\subcaption{The Icicle workflow. Filesystem usage information is monitored using the Octopus MCP server and user data are compared and visualized from the Globus Search MCP server.}
\label{fig:icicle-app}
\end{minipage}

\caption{The four application scenarios. In each, the Claude Desktop agent translates a user-supplied prompt into an agentic application that invokes MCP tools. The subfigures show where those tools perform their actions.}
\label{fig:four_panel}
\end{figure*}

We employ four use cases, shown in~\autoref{fig:four_panel}, to explore how our MCP-oriented architecture applies to scientific workflow development. 
In each case, we use Claude Desktop as the agentic assistant and Claude Sonnet 4 as the LLM. (We have also experimented with other assistants and LLMs, but not yet in any systematic manner.) We present the user prompts for each use case in the Appendix.

\subsection{Molecular Structure with Garden}

We evaluate the Garden MCP server through an end-to-end use case that demonstrates the full machine learning workflow cycle, from rapid prototyping to production-scale execution. The application involves molecular structure relaxation using machine-learned interatomic potentials (MLIPs), progressing from single-structure exploration to batch processing of multiple structures.

The two-part \cref{prompt:garden1} defines an agentic application that first uses the Garden MCP server to discover a collection of MLIP models, from which it selects a MACE~\cite{Batatia2022mace} model variant for rapid prototyping. Given local user data, specified in the prompt, describing a single 32-atom copper structure, the agent manages data staging and then optimizes the structure to its lowest energy state (-130.71 eV) with typical FCC copper lattice parameters. It then leverages the Garden MCP server to relax the structure remotely and returns the results with a brief analysis for the user, enabling quick exploration of ML models without local dependency management. The application then scales to production execution, batch processing 49 copper structures with the same MACE model on ALCF's Edith cluster using standard HPC patterns for job submission, status polling, and results retrieval.

\subsection{Multi-site Phylogenetic Analysis}

Computational biology can benefit from workflows that leverage 
specialized computing resources, but researchers face 
challenges coordinating data and analyses across distributed HPC systems. 

We perform this task via an agentic application that employs Globus MCP servers to run a multi-method phylogenetic analysis workflow using FastTree~\cite{price2010fasttree}, RAxML~\cite{Stamatakis2014RAxML}, and IQ-TREE~\cite{Minh2020IQtree} to validate evolutionary relationships among motor proteins.
The application uses Globus MCP servers to distribute tasks across ALCF Polaris, for data acquisition and sequence alignment, and NERSC Perlmutter, for computationally intensive RAxML phylogenetic reconstruction with bootstrap analysis. 
The MCP servers handle authentication, job submission, status monitoring, and data transfer between ALCF and NERSC systems, abstracting the complexity of multi-site resource management.  

\Cref{prompt:phylo1}
provides context about relevant Globus infrastructure, including Transfer and Compute endpoints for both ALCF Polaris and NERSC Perlmutter; their configured execution environments; and available bioinformatics software. 
It also specifies target bacterial species for evolutionary analysis and requests that the agent verify that Polaris and Perlmutter were available, then download protein sequence accessions to Polaris, perform sequence alignment, replicate the aligned data to NERSC, execute phylogenetic reconstruction using different algorithms across both sites for comparative validation, and finally, return results.

The agent runs the workflow by dynamically creating Python functions for the necessary analysis steps and correctly passing inputs between tools.
This approach eliminates the need for researchers to develop custom glue code between different computational tools or to adapt their code to specific HPC systems, as the agent autonomously handles integration complexities.

\subsection{Quantum Chemistry}
 
Workflows in this domain typically require that researchers manually orchestrate sequences of quantum chemistry calculations---work that can require significant software integration effort. The computational expertise needed to translate conceptual questions into executable code creates bottlenecks that slow exploration. 

We evaluate our MCP servers using a quantum chemistry workflow that demonstrates automated computational pipeline generation for materials science applications. This approach uses quantum chemistry computational codes to provide computation-backed results rather than relying on potentially inaccurate estimates from LLMs.
We configured an environment on ALCF's Polaris with various quantum chemistry software, including PySCF~\cite{sun2018pyscf} and GPU4PySCF~\cite{li2025introducing}. We equip an agent with the Globus Compute MCP server to orchestrate the remote execution of code. 

We use~\cref{prompt:chem1} to instruct an agent to perform Highest Occupied Molecular Orbital (HOMO) – Lowest Unoccupied Molecular Orbital (LUMO) gap calculations~\cite{griffith1957ligand} for six common organic solvents used in battery electrolytes. HOMO-LUMO gaps provide valuable information for determining electrochemical stability windows when designing battery electrolytes. 
This calculation involves multiple steps that traditionally require significant chemistry expertise: guessing initial molecular structures, optimizing structures to stable states, calculating orbital energies, and determining HOMO-LUMO energy differences with proper software inputs and parameter selection. 
Using the Globus Compute MCP and quantum chemistry software, the agent wrote and registered the necessary Python functions for execution via Globus Compute on Polaris, monitored task status, and upon completion, generated visualizations of the calculated HOMO-LUMO gaps.
The computed HOMO-LUMO gaps (see \autoref{fig:homo-lumo-gap})
are consistent with published literature~\cite{shakourian2016evaluating}, confirming that the agent correctly executed this complex quantum chemistry computation and demonstrating the potential for this capability to accelerate more sophisticated computational chemistry research.

\subsection{Filesystem Monitoring}

In this fourth use case, we consider the problem of understanding user behavior and system performance in complex storage systems, via synthesis of data from multiple monitoring tools and time scales. 

The Icicle application~\cite{icicle} leverages the MCP architecture for real-time HPC filesystem monitoring and analytics. In this use case, we deployed the Icicle monitoring software on a Lustre filesystem to continuously monitor filesystem events and report them through the Octopus event fabric. These events are then processed and published into a Globus Search index for storage and exploration. 

\Cref{prompt:icicle1} engages the agent to use the Octopus MCP server to evaluate recent filesystem activity by querying an Octopus topic and retrieving information about filesystem events over the preceding hour, providing insight into system utilization patterns and potential performance bottlenecks.
The workflow then uses the Globus Search MCP server to generate data summaries for individual users by querying the search index and producing a report and visualization of their data usage, including file count, average file size, and total volume used across the filesystem. 

This dual-phase approach enables administrators to correlate system-wide activity trends with individual user behaviors, facilitating informed decisions about resource allocation, storage optimization, and system maintenance scheduling.

\section{Discussion}

MCP presents a promising architecture for providing agents with access to research CI and enabling
agents to orchestrate complex workflows across heterogeneous systems that span different administrative domains, authentication systems, and computing platforms. 
In our use cases, agents equipped with scientific MCP servers demonstrate a remarkable ability to recover from failures, adapt to changing resource availability, and dynamically adjust strategies when initial approaches encounter problemsadding flexibility to traditionally rigid systems while also lowering barriers to entry.

\subsection{Dynamic Tool Discovery}

\Rhea~demonstrates how MCP servers can overcome the challenges associated with exposing thousands of tools by replacing static registries with retrieval-based discovery. Vectorizing large collections of tool descriptions and applying semantic similarity search is especially valuable when the query and tool descriptions differ in terminology or level of abstraction. 
The approach may be even more valuable in scenarios where the search corpus extends beyond static metadata to include execution logs, user feedback, and runtime metrics, enabling discovery informed by both descriptions and historical performance in relevant contexts. 
Such a mechanism may enable each user query to contribute to and benefit from the collective knowledge embedded in the tool repository, fostering an extensible ecosystem that grows more capable as new scientific tools are added.

To evaluate \Rhea’s approach, we used \texttt{Llama-3.3-70B} to generate a benchmark of 380 queries derived from Galaxy training tutorials~\cite{galaxy-training}. Each tutorial defines one or more workflows, where each workflow specifies the sequence of Galaxy tools required to complete the analysis task. We used these workflows to establish ground-truth mappings between natural-language problem descriptions and the corresponding tools. For example, from the ``proteogenomics-1-database-creation'' tutorial~\cite{PGtut}, we derived the query ``I need a tool to compare and evaluate the accuracy of RNA-Seq transcript assemblers,'' which maps to the Galaxy tool \texttt{gffcompare}~\cite{pertea2020gff}. 
This evaluation design enables direct measurement of retrieval precision against real training materials.

\begin{figure}[t!]
    \centering
    \includegraphics[width=1\linewidth, trim={0.2cm, 0.4cm, 0.3cm, 1.0cm}, clip]{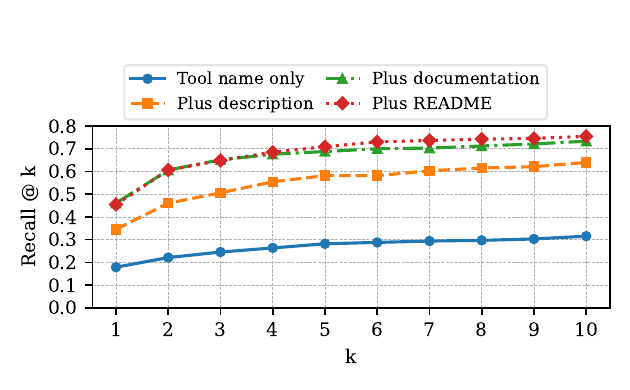}
    \caption{\Rhea's Galaxy Tool retrieval performance (Recall@k) across four documentation embedding strategies.
    }
    \label{fig:rhea-recal}    
\end{figure}

\Cref{fig:rhea-recal} reports \Rhea’s retrieval performance as Recall@k across four documentation embedding strategies. Each query maps to a single ground-truth tool; Recall@k reflects whether the correct tool appears within the top-$k$ retrieved results, averaged over 380 queries. Incorporating richer textual context improves performance: embeddings based solely on tool names perform worst, while adding descriptions, extended documentation, and repository README files progressively enhances retrieval accuracy.

\subsection{Authentication}

Our implementation revealed important lessons regarding authentication and authorization within the MCP framework, particularly when integrating with Globus services that rely on OAuth-based authorization protocols.
For example, handling OAuth token management and session persistence directly within hosted MCP server deployments adds complexity associated with secure token passing, token refresh cycles, and iterative authentication and scope management.
To avoid these challenges, we instead operate the MCP servers locally within the user's trusted environment and wrap all Globus service interactions with an authentication handler that dynamically manages authentication flows and acquires additional scopes as needed.
This architecture simplifies credential management, eliminates session synchronization issues between remote servers and OAuth providers, and provides users with direct control over their authentication flows while maintaining the security and authorization benefits of the Globus ecosystem.

As the authentication scheme for MCP is rapidly evolving, we will continue our investigation by evaluating emerging methods like fastMCP's new Remote OAuth capabilities~\cite{remote-oauth}. This feature may offer new solutions for managing distributed authentication in hosted MCP server environments.

\subsection{Resilience}

Our experiences with the MCP-oriented architecture revealed surprising capabilities in agent self-correction when encountering errors during scientific workflows. We observed agents successfully diagnosing and recovering from various failure modes, such as when a Globus Compute function failed and the agent processed the resulting Python error to determine that an analysis invocation was malformed, then automatically reformulated the request. Similarly, agents would submit Globus Transfer tasks with incorrect paths. When a task failed, the agent processed the error returned when checking task status and then resubmitted the transfer with the correct path. This autonomous error handling demonstrated the potential for resilient scientific computing workflows that can adapt to common failure scenarios without human intervention.

Our evaluation also revealed limitations in agent resilience patterns. Agents made repetitive mistakes, suggesting limited learning from previous errors within the same session. Furthermore, they would not always complete outlined tasks as specified, or would produce inconsistent outputs such as different visualizations for similar requests, indicating variability in task interpretation and execution. These findings highlight the need for improved resilience mechanisms for agent-driven scientific workflows, including techniques for learning from past failures, maintaining consistency across similar tasks, and implementing systematic approaches to error recovery and workflow completion verification.

\section{Conclusion}
We have presented MCP servers for scientific research and evaluated their effectiveness in four scientific use cases.
These implementations show how the MCP architecture can reduce barriers to using complex and distributed research CI.
We discussed the advantages of building thin MCP adapters for existing research services and separating discovery from invocation to manage large tool ecosystems, and showed how agents can resiliently and dynamically generate the glue code needed for scientific workloads. Open challenges include hosting cross-domain MCP services under current authentication models, evaluating agent workflow reliability, improving resilience for long-running tasks, and evaluating MCP use with different agents and LLMs.

\balance
\bibliographystyle{ACM-Reference-Format}
\bibliography{refs/refs}

\twocolumn
\appendix
\nobalance

\begin{table*}[t]
\centering
\caption{Characteristics of the four MCP-based scientific applications considered in this article.}
\label{tab:mcp_applications}
\begin{tabular}{|l|c|c|c|c|}
\hline
& \textbf{Phylogenetic} & \textbf{Molecular Design} & \textbf{Quantum Chemistry} & \textbf{Icicle} \\
\hline
\textbf{LLM ($\mathcal{L}$)} & Claude Sonnet 4 & Claude Sonnet 4 & Claude Sonnet 4 & Claude Sonnet 4 \\
\hline
\textbf{Agent Platform} & Claude Desktop & Claude Desktop & Claude Desktop & Claude Desktop\\
\hline
\textbf{MCP Servers ($\mathcal{M}$)} & 
\begin{tabular}[c]{@{}c@{}}Globus Compute\\Globus Transfer\\Facility Status\end{tabular} & 
\begin{tabular}[c]{@{}c@{}}Garden\\Globus Compute\end{tabular} & 
\begin{tabular}[c]{@{}c@{}}Globus Compute\end{tabular} &
\begin{tabular}[c]{@{}c@{}}Octopus\\Globus Search\end{tabular} \\
\hline
\textbf{Sites ($\mathcal{S}$)} & ALCF, NERSC & ALCF, Cloud & ALCF & Local \\
\hline
\textbf{Key Software ($\Pi$)} & 
\begin{tabular}[c]{@{}c@{}}FastTree,\\RAxML, IQ-TREE\end{tabular} & 
\begin{tabular}[c]{@{}c@{}}MACE,\\ASE\end{tabular} & 
\begin{tabular}[c]{@{}c@{}}PySCF,\\GPU4PySCF\end{tabular} & N/A \\
\hline
\end{tabular}
\end{table*}

\section*{Appendix}
\label{sec:appendix}
\crefalias{section}{appendix}
\counterwithin{figure}{section}
\setcounter{figure}{0}

The appendices provide supplementary details for the applications presented in the main text, including the initial prompts, additional explanations, analysis results, and their visualization.
For a summary of LLM, agent platform, MCP servers, sites, and software involved, see \Cref{tab:mcp_applications}.

\section{Molecular Structure with Garden}\label{app:garden}

This application showcases a complete molecular materials science workflow, from rapid prototyping to production-scale execution. We use MACE for molecular structure relaxation, first testing on a single structure before scaling up to a batch of 49 structures on ALCF's Edith HPC cluster. \Cref{prompt:garden1} shows the initial prompt given to Claude Desktop (with Claude Sonnet 4) to launch the workflow.

\begin{promptbox}[label=prompt:garden1]
I have an XYZ structure file at <USER-DIRECTORY>/single-material.xyz that I’d like to relax first using a suitable MLIP model on the cloud. 
Please optimize the structure to its lowest energy state, save the relaxed structure to <USER-DIRECTORY>/single-material-relaxed.xyz and provide a brief summary of the result, including the final energy and lattice parameters. 
\\\\
After that, I’d like to run a batch relaxation job on ALCF’s Edith cluster using another XYZ file at <USER-DIRECTORY>/test-structures.xyz. 
Use the same MACE MLIP model to relax all structures in this file, submit the job to Edith, monitor its status, and, once complete, save the results to <USER-DIRECTORY>/batch-relaxed-structures.xyz and provide an overall summary.\\
{\normalfont [Two XYZ structure files are attached.]}
\end{promptbox}

\begin{figure}[t!]
    \centering
    \includegraphics[width=\linewidth,]{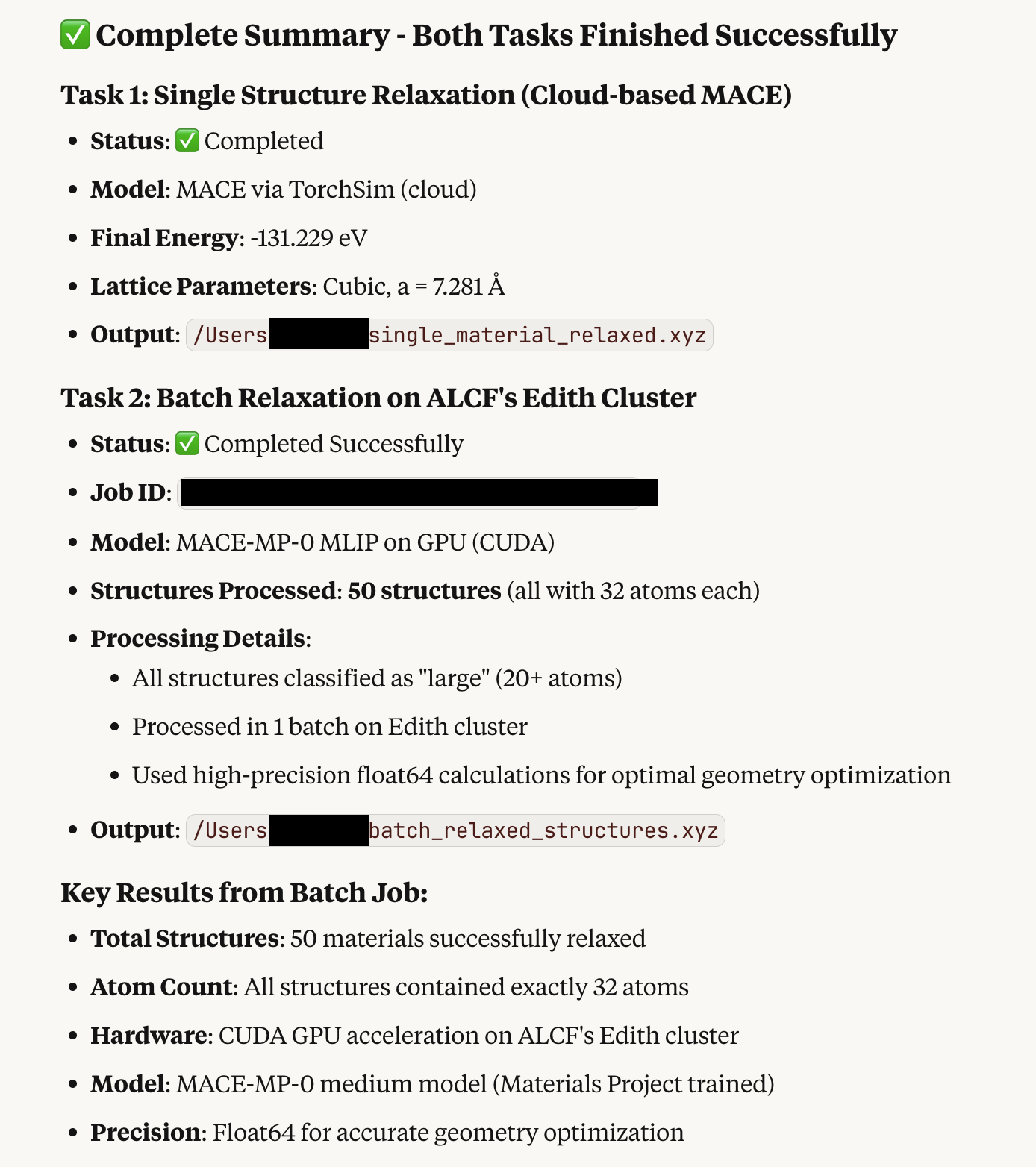}
    \caption{A redacted screenshot of the final output from the Garden workflow.}
    \label{fig:garden-visuals}
\end{figure}

As shown in the redacted screenshot in \Cref{fig:garden-visuals}, the final output panel lists the models chosen for each task, provides storage paths for the results, and summarizes key outcomes like the number of relaxed structures, hardware used, and computational precision.

\section{Multi-site Phylogenetic Analysis}\label{app:phylo}

This use case performs a phylogenetic analysis workflow that coordinates computations between ALCF Polaris and NERSC Perlmutter. To explore the evolutionary relationships among bacterial motor proteins, the workflow downloads data, performs a sequence alignment, and runs three different phylogenetic reconstruction tools (FastTree, RAxML, and IQ-TREE) across the sites for validation. \Cref{prompt:phylo1} supplied Claude Desktop with the necessary infrastructure details, experiment steps, and scientific goals.

\begin{promptbox}[label=prompt:phylo1]
Consider the following.
\\\
ALCF Polaris:\\
- Compute endpoint: ...\\
- Transfer endpoint: ...\\
- Working dir: <USER-DIRECTORY>/working\\
- Results dir: <USER-DIRECTORY>/results\\
NERSC Perlmutter:\\
- Compute endpoint: ...\\
- Transfer endpoint: ...\\
- Working dir: <USER-DIRECTORY>/working\\
\\\
Each Globus Compute endpoint has the following packages installed:  
biopython, bioconda, mafft, clustalo, muscle, fasttree, raxml, iqtree.
Work within the specified working directories. 
When transferring data from ALCF, replace /eagle/ with / for the input path as the endpoint is mounted at /eagle/.\\
\\\
Execute a workflow that will perform the following tasks on Polaris and NERSC.

\begin{enumerate}[leftmargin=2em, label=\arabic*.]
\item Check ALCF and NERSC are both online.
\item At Polaris download a text file of accessions (one per line) or pipe the output of NCBI ESearch queries such as fliC[gene] AND "Escherichia coli"[orgn] AND refseq[filter]. Set "Return type" to "Protein fasta". Use dataset collections to keep one FASTA per species. Set retmax=1 on the esearch query.
\item Merge the FASTA datasets into one collection on Polaris and then run a sequence alignment. Typical settings: default gap penalties; for MAFFT you may choose "L-INS-i" for highest accuracy. Choose one of mafft, muscle, or clustalo
\item Transfer the aligned file to NERSC’s working directory. Remember to specify the filename in the destination path.
\item Build phylogenetic trees (using all three of FastTree, RAxML, IQ-TREE). At Polaris, use for IQ-TREE 2 leave "Auto" model selection on and request 1 000 ultrafast bootstraps and FastTree with JTT+CAT (protein). 
\item At NERSC, use RAxML: PROTGAMMAJTT and 100 rapid bootstraps.  Run these tasks in parallel.
\item Move the NERSC output back to the Polaris results directory.
\item Finally, retrieve the results and plot and visually compare the phylogenetic trees side-by-side, ensuring consistent layout and labelling for easy comparison of topologies and branch lengths.
\end{enumerate}
\noindent
Use the following species:
\begin{verbatim}
[
  ("fliC", "Escherichia coli"),
  ("fliC", "Pseudomonas aeruginosa"),
  ("flaB", "Borrelia burgdorferi"),
  ("fliC", "Serratia marcescens"),
  ("fliC", "Shewanella oneidensis"),
  ("flaA", "Vibrio cholerae"),
  ("fliM", "Listeria monocytogenes"),
  ("fliC", "Salmonella enterica"),
  ("flaA", "Agrobacterium tumefaciens"),
  ("fliC", "Clostridioides difficile"),
]
\end{verbatim}
\noindent
Perform the tasks as individual Compute functions where possible and work in the ~/testing/ directory and write logs and output from each step to this directory. Once a task is submitted periodically poll for its status until it is completed. Wait one minute between polling task status.

\end{promptbox}

\begin{figure}[t]
    \centering
        \centering
        \includegraphics[width=\linewidth,]{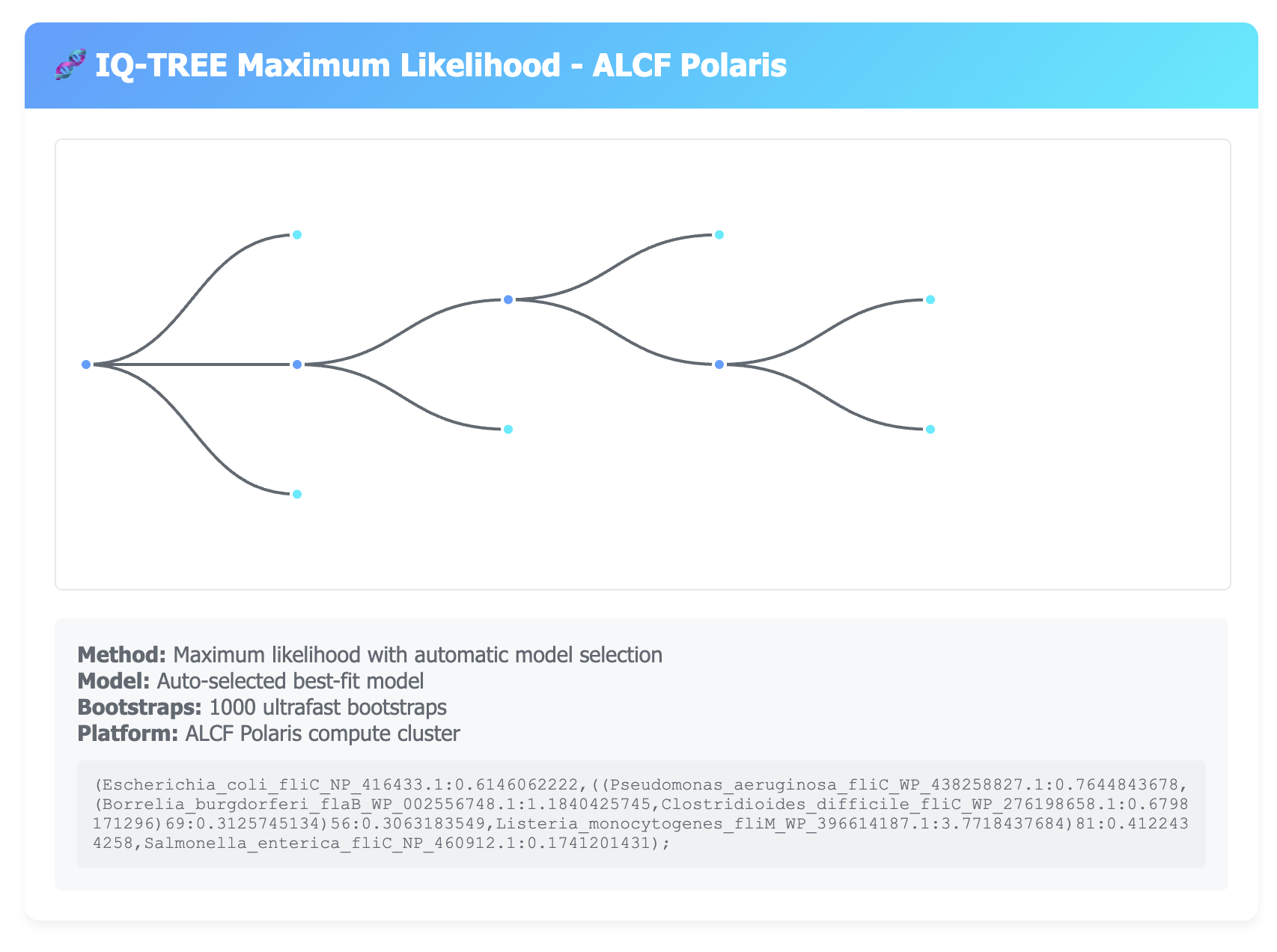}
        \caption{Tree reconstruction using IQ-TREE on ALCF Polaris from the phylogenetic workflow.}
        \label{fig:phylo-1}
    \hfill
        \centering
        \includegraphics[width=0.99\linewidth,]{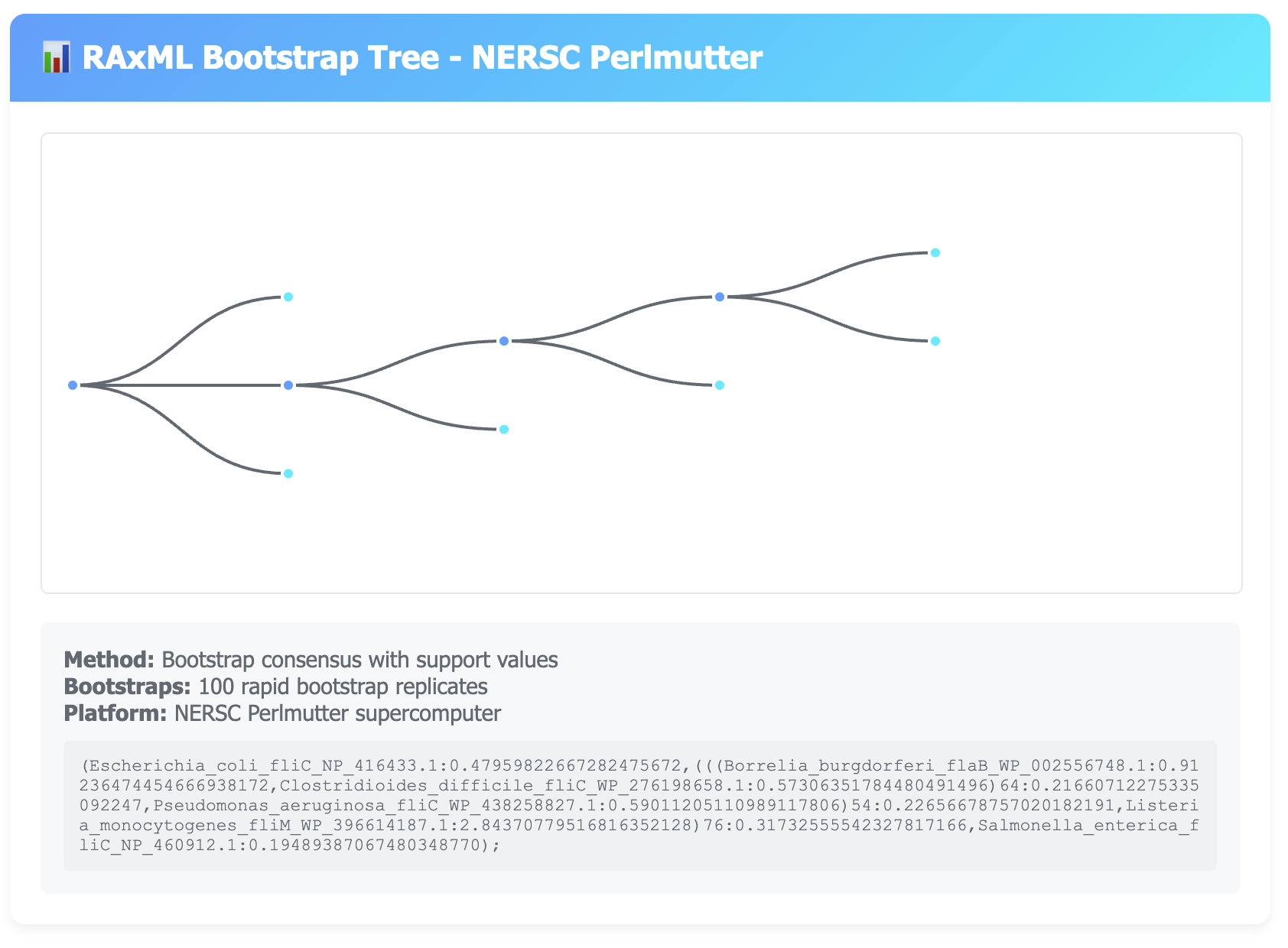}
        \caption{Tree reconstruction using RAxML on NERSC Perlmutter from the phylogenetic workflow.}
        \label{fig:phylo-2}
    \label{fig:phylo-visuals}
\end{figure}

The final output is a comprehensive summary presented in four sections. It begins with an ``Analysis Summary'' of key metrics and a list of the six ``Analyzed Species and Genes''. The main results are presented under ``Tree Reconstructions,'' which displays the five trees generated by the three methods (i.e., FastTree, RAxML, IQ-TREE); two examples, IQ-TREE on ALCF Polaris and RAxML on NERSC Perlmutter, are shown in \Cref{fig:phylo-1} and \Cref{fig:phylo-2}, respectively. The output concludes with ``Phylogenetic Analysis Results,'' which provides some key observations and technical notes as paragraphs.

\section{Quantum Chemistry}\label{app:chem}

\begin{figure}[ht]
    \centering
        \centering
        \includegraphics[width=\linewidth, page=1, trim={0, 3cm, 10cm, 0cm}, clip]{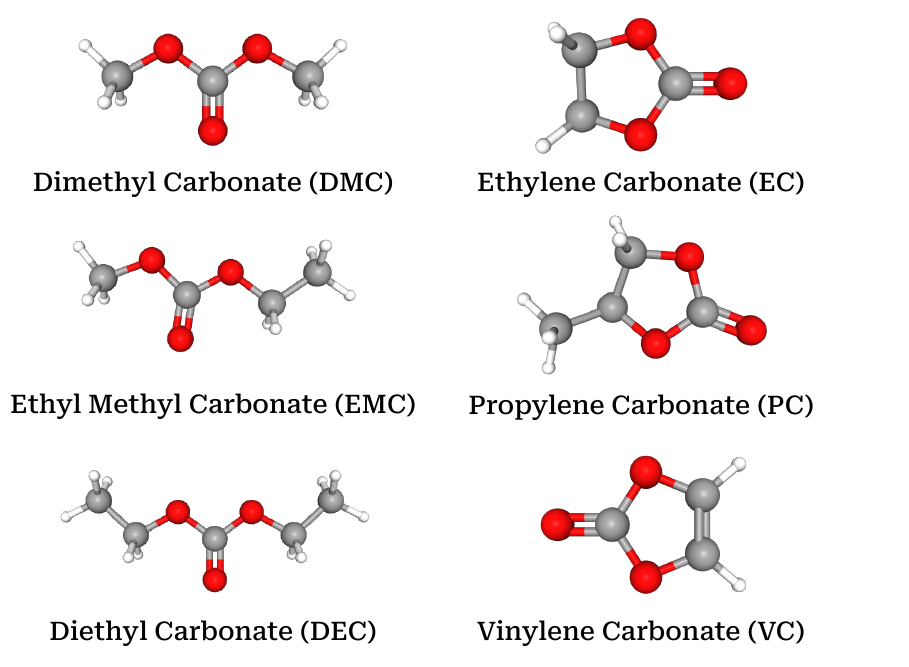}
        \caption{Molecular structures of the six carbonate solvents used in the quantum chemistry application: DMC, EMC, DEC, EC, PC, and VC. Carbon is gray, oxygen red, and hydrogen white.}
        \label{fig:molecules-struct}
    \hfill
        \centering
        \includegraphics[width=\linewidth, trim={0.2cm, 0.4cm, 0cm, 0cm}, clip]{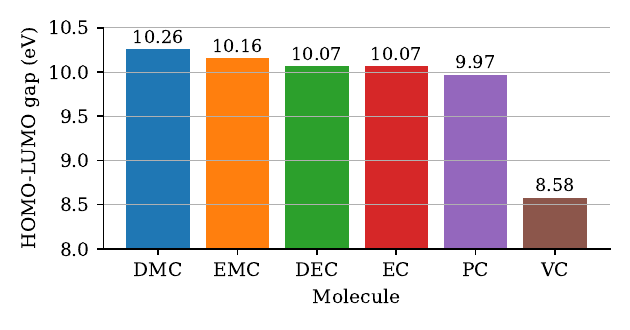}
        \caption{Computed HOMO-LUMO gaps from GPU4PySCF calculations at the M06-2X/6-311++G(d,p) level using the Globus Compute MCP server on Polaris.}
        \label{fig:homo-lumo-gap}
    \label{fig:molecules}
\end{figure}

This application calculates the HOMO-LUMO gap, a key property for designing stable battery electrolytes. We investigate six common carbonate solvents: three linear (DMC, EMC, DEC) and three cyclic (EC, PC, VC), as shown in \autoref{fig:molecules-struct}. Although all share a core carbonate functional group \mbox{–O–C(=O)–O–}, their differing structures and substituents influence their electrochemical behavior, motivating this comparative study.

To automate this task, we provided Claude Desktop with the high-level prompt shown in \cref{prompt:chem1}. Claude Desktop orchestrated the entire two-step workflow—structure optimization followed by orbital energy calculation—by generating the necessary Python functions. It then executed these functions using the GPU4PySCF software on ALCF Polaris, managing the process via the Globus Compute MCP server and returning the final results for analysis. The calculated HOMO-LUMO gaps are shown in \Cref{fig:homo-lumo-gap}.

\begin{promptbox}[label=prompt:chem1]
    I have a quantum chemistry computation globus-compute endpoint available at ... It has the GPU4PySCF package. 
    \\\\
    Could you use GPU4PySCF to optimize the structure and calculate the HOMO-LUMO gap of these molecules at M06-2X/6-311++G(d,p) level? 
    \\\\
    EC (Ethylene Carbonate), PC (Propylene Carbonate), VC (Vinylene Carbonate), DMC (Dimethyl Carbonate), EMC (Ethyl Methyl Carbonate), and DEC (Diethyl Carbonate). 
    Their initial structures are included in these files.\\ 
    {\normalfont [Six XYZ structure files are also attached.]}
\end{promptbox}

\section{File System Monitoring}\label{app:icicle}
This use case applies the MCP architecture to an HPC administration task: real-time filesystem monitoring and analytics. Using the Icicle monitoring software, filesystem events are published to the Octopus event fabric and indexed in Globus Search. The workflow first uses the Octopus MCP server to get real-time changelogs of system activity and then uses the Globus Search MCP server to perform a historical analysis and comparison of specific users' storage patterns. The prompt shown in \cref{prompt:icicle1} instructs Claude Desktop to perform this two-phase analysis.

\begin{promptbox}[label=prompt:icicle1]
    Consume Icicle changelogs from the lustre-mon-out topic on the Octopus event fabric to summarize today’s usage. Retrieve up to 10,000 events within a 10-second timeout, providing a concise overview of notable activity patterns and key insights. Do not downsample during analysis.
    \\\\
    Next, query the Globus Search index ... for records with the subjects ``user\_id::<USER-A>'' and ``user\_id::<USER-B>''. Perform a comprehensive comparison of these two users across all available dimensions.
    \\\\
    Visualize results using lightweight, token-efficient tables and graphs. Ensure all identifiers are anonymized and obfuscated, except for labeling the two users as UserA and UserB.
\end{promptbox}

\begin{figure}[t!]
    \centering
        \centering
        \includegraphics[width=\linewidth,]{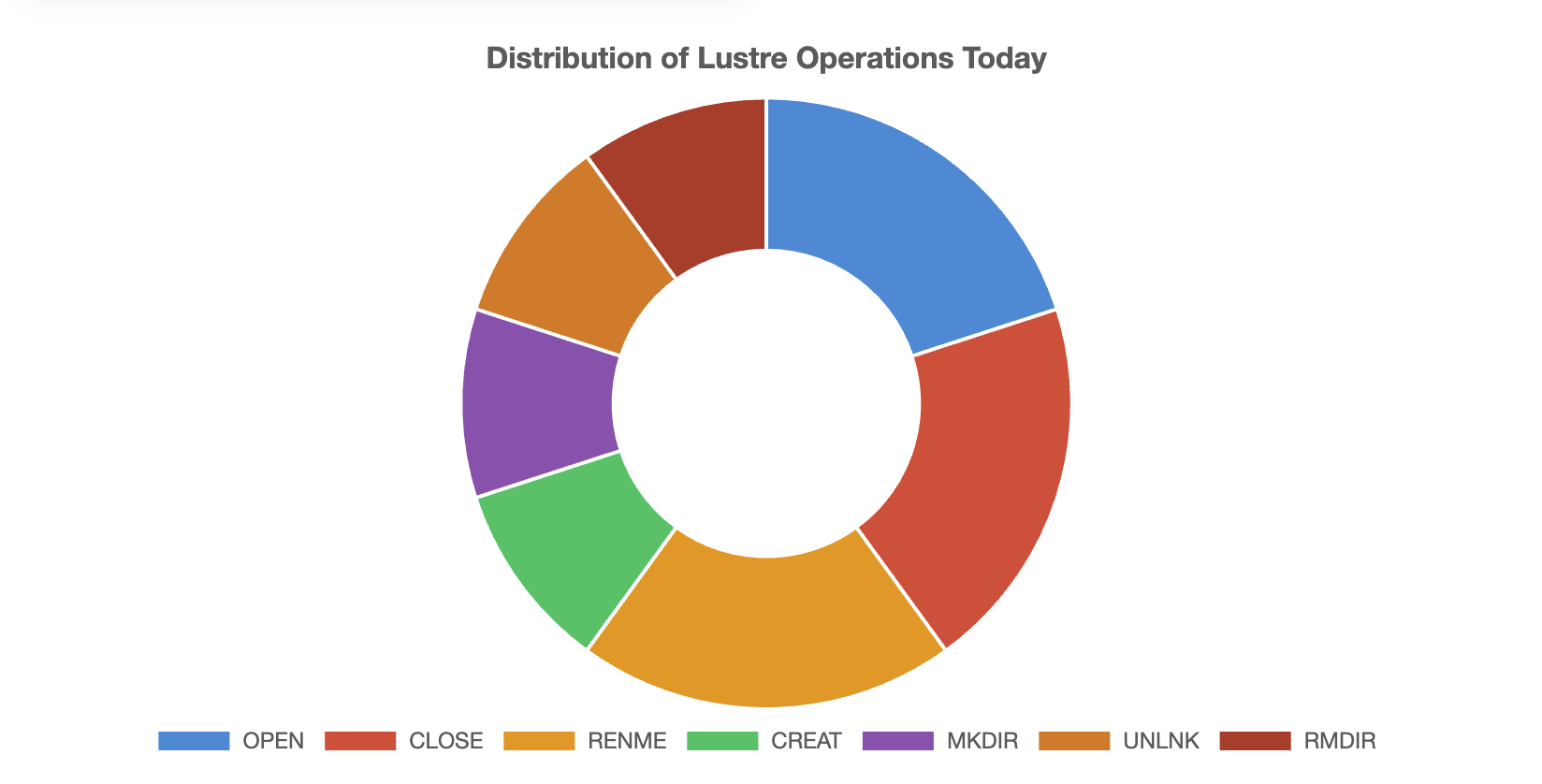}
        \caption{Distribution of Lustre filesystem events. This summary is based on 87 events sampled by the agent from the Octopus event fabric. Each segment represents an event type (e.g., CREAT, OPEN, UNLNK).}
        \label{fig:icicle-1}
    \hfill
        \centering
        \includegraphics[width=\linewidth,]{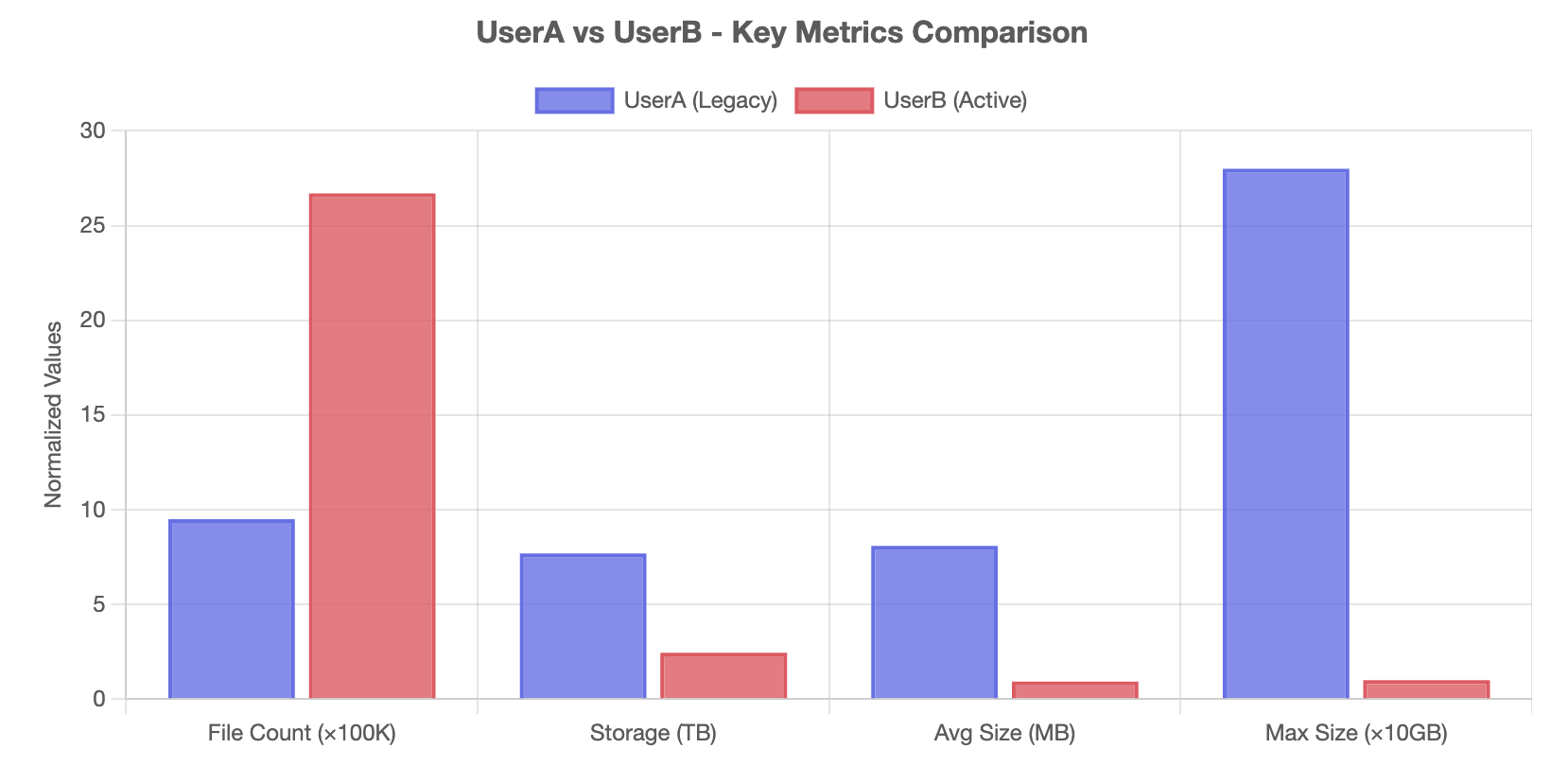}
        \caption{Comparison of storage metrics for the two users. The chart compares across total file count (in hundreds of thousands), total storage used (TB), average file size (MB), and maximum file size (GB).}
        \label{fig:icicle-2}
    \label{fig:icicle-visuals}
\end{figure}

In response to the prompt, Claude Desktop generated a two-part visualization panel.

The first panel, titled ``Today's Lustre Activity Summary,'' presents a pie chart about the event type distribution, as shown in \autoref{fig:icicle-1}. The visualization is interactive: clicking a legend removes the corresponding section on the ring. From sampled changelogs, Claude Desktop inferred several key activity patterns, summarized in three paragraphs, titled ``Repetitive Workflow,'' ``Automated Process,'' and ``Resource Usage,'' respectively. This panel concludes with a table detailing the sampled changelog types in four columns: ``Operation Type'', ``Count'', ``Percentage'', and ``Description''.

The second panel, ``User Activity Comparison: UserA vs. UserB,'' provides a comparative analysis of two users based on data queried from Globus Search. It begins with a high-level summary comparing metrics such as  ``Total Files,'' ``Storage Used,'' ``Avg File Size,'' ``Activity Period,'' and ``Peak Activity.'' This is complemented by a comparative bar chart, shown in \autoref{fig:icicle-2}, which visualizes four attributes: total file count, total storage used, average file size, and maximum file size. Note that the summary and the bar chart have three attributes in common. Hovering the mouse over bars reveals the underlying numerical data. Claude Desktop also provides three paragraphs of behavioral analysis based on the queried data, titled ``UserA,'' ``UserB,'' and ``Efficiency Insight'' respectively.  
Finally, it concludes with another summary table of attributes as rows: ``Total Storage (TB)," ``Avg File Size (MB),'' ``Max File Size (GB),'' ``Median File Size (KB).'' All except the median file size have appeared in previous visualizations. The table has an additional column of a calculated ratio of UserB's metrics relative to UserA's, offering an intuitive comparison.

\clearpage 
\end{document}